\documentclass[letterpaper,conference]{IEEEtran}
\usepackage[utf8]{inputenc}
\usepackage{url}
\usepackage{hyperref}
\hypersetup{
  pdfinfo={
    Title={A Proposal for the Measurement and Documentation of Research Software Sustainability in Interactive Metadata Repositories},
    Author={Stephan Druskat},
    Subject={Research software sustainability},
    Keywords={Metadata, repositories, technical sustainability, information search and retrieval, measurement, software management plans}
  }
}
\usepackage{paralist}
\usepackage{enumitem}
\setlist{nosep}

\IEEEoverridecommandlockouts
\IEEEpubid{\makebox[\columnwidth]{This work is licensed under a \href{https://creativecommons.org/licenses/by/4.0/}{CC-BY-4.0 license}. \hfill} \hspace{\columnsep}\makebox[\columnwidth]{ }}

\title{A Proposal for the Measurement and Documentation of Research Software Sustainability in Interactive Metadata Repositories}
\author{\IEEEauthorblockN{Stephan Druskat}
\IEEEauthorblockA{Department of German Studies and Linguistics\\
Humboldt-Universität zu Berlin\\
Berlin, Germany\\
stephan.druskat@hu-berlin.de}}

\begin{document}

\maketitle

\begin{abstract}
	This paper proposes an interactive repository type for research software metadata which measures and documents software sustainability by accumulating metadata, and computing sustainability metrics over them. Such a repository would help to overcome technical barriers to software sustainability by furthering the discovery and identification of sustainable software, thereby also facilitating documentation of research software within the framework of software management plans.
\end{abstract}

	The identification and discovery of sustainable research software present technical barriers to software sustainability \cite[p. 16f.]{dfg-knowledge-exchange}. This paper proposes to tackle these barriers by accumulating and quantifying software metadata in interactive metadata repositories (IMRs) to enable identification of sustainable software through \emph{sustainability measurement}, and simplified discovery of suitable software through \emph{documentation of software and its sustainability}.

	In analogy to the discussion about data sustainability and funders' mandate of data management plans (DMPs), software management plans (SMPs) would be an obvious tool for securing sustainability of newly developed software. DMPs feature data repositories as natural targets for research data releases as part of the data lifecycle documentation, and analogously, software repositories should be an integral part of software lifecycles as detailed in SMPs. In addition to source code repositories and distribution repositories for deliverables, there are also already a large number of research data repositories which archive research software source code and deliverables along with some metadata (cf. \href{http://re3data.org/}{re3data.org}). However, while the latter may provide support for assessing the suitability of a software for the intended research application, they do not provide a lot of information about, not to mention metrics of, the software's sustainability. The same is true for existing \emph{metadata} repositories dedicated to research software, e.g., \href{http://sciencepad.web.cern.ch/}{SciencePAD}, the \href{https://www.egi.eu/services/catalogue/appdb.html}{EGI Applications Database}, and the \href{http://dirtdirectory.org/}{DiRT Directory}. There are also other platforms providing metadata for software, e.g., \href{https://github.com/}{GitHub}, \href{https://www.openhub.net/}{Open Hub}, and \href{https://zenodo.org/}{Zenodo}. Some of the metadata they provide -- a rather small portion in the case of GitHub and Zenodo, some more in the case of Open Hub -- can indeed be used to discover some sustainability aspects of the software they cover. However, this facet is not presented explicitly in either of the platforms. In any case, considering the available tools and platforms, users that are interested in acquiring a comprehensive and in-depth understanding of how sustainable a software is will have to interpret and extrapolate to a high degree. 

	However, a repository can be a suitable means to provide documentation and measurement of software sustainability, specifically by utilising software metadata. In order to do so, it must: accumulate the metadata for a software that is relevant for determining both its suitability for a research application (to allow for straightforward discovery of software), and its sustainability; compute sustainability metrics; document the metrics and provide means to interactively evaluate and refine some of them. Such a repository would also benefit projects which will be subject to funding agencies' potential mandate of SMPs, as it represents a natural means of documenting the sustainability of software developed within the project.

	The conception of the envisioned IMR poses at least three theoretical challenges, namely (1) how to give ``software sustainability'' an operationalisable definition, (2) how to define quantifiable criteria for sustainability, and (3) how to design algorithms for computing comprehensible and reproducible metrics for sustainability. While detailing (3) is beyond the scope of this paper and is left for future research, (1) and (2) are approached briefly in the following sections.

	``Software sustainability'' is an ambiguous concept, with different agents approaching it from different angles. Reference \cite{umweltbundesamt} focuses on the ecological sustainability of software, discussing software engineering only briefly. Reference \cite{agile-sustainable-software-development} focuses on the development process rather than the product. Reference \cite{better-software} -- specifically discussing research software -- names training, availability, community recognition for developers, dedicated job descriptions, and funding as defining factors for sustainability. A systematic approach to a definition of sustainability in the context of software engineering is presented in \cite{penzenstadler-definition-sustainability}, which utilizes a basic definition of sustainability as ``preserving the function of a system over a defined timespan'' \cite[p. 1184]{penzenstadler-definition-sustainability} for the refinement of detailed aspects of sustainability for and in software engineering. However, while \cite{penzenstadler-definition-sustainability}'s approach would be operationalisable, it uses a generic, multi-dimensional definition of sustainability (cf. \cite[p. 4]{generic-model-sustainability}) and applies it to \emph{engineering processes} rather than to \emph{features of products} and is therefore not directly adaptable for IMRs. 

	A definition of sustainability that can be operationalized for the purposes of an IMR must in fact be based on the definition of ``technical sustainability'' from \cite[p. 470f.]{karlskrona-manifesto-paper}: ``Technical [sustainability] refers to longevity of information, systems, and infrastructure and their adequate evolution with changing surrounding conditions.'' The two key points of this definition, \emph{longevity} and \emph{evolution}, can be concretized by transforming another existing global concept of ``sustainability''. Following \cite{kit-nachhaltigkeit}, an enhancement of a three-dimensional general model of sustainability (``three column model''), we can preliminarily define ``software sustainability'' via its goals: \emph{The three goals of software sustainability are}
	\begin{inparaenum}[(1)]
		\item \emph{ensuring the existence of the software,}
		\item \emph{preserving the potential for productive operation of the software,}
		\item \emph{creating and retaining possibilities for further development and adaptation of the software.}
	\end{inparaenum}
	This definition, finally, should be employable for defining and categorizing suitable criteria for measuring the sustainability of research software.

	An IMR must accumulate quantifiable metadata pertaining to all three sustainability goals. The parameters for the metrics to be computed over these metadata must be based on criteria that are themselves categorized along the lines of the sustainability goals. One starting point for the compilation of a list of criteria is \cite{software-evaluation-criteria}, which defines criteria for ``sustainability'', ``maintainability'', and ``usability'' to facilitate the evaluation of software products (e.g.: ``Documentation is on the project web site'', ``software has an open source licence'', ``source code is commented''). As the definition for sustainability given above includes all three of these notions (usability arguably counting towards goal (2)), the items given for the criteria in \cite{software-evaluation-criteria} can be transformed into parameters for computing sustainability metrics for an IMR, although some items may require further definitory work to make them quantifiable. Further criteria that are not included in \cite{software-evaluation-criteria} -- e.g., pertaining to human resources available for a software project -- can be identified introspectively, elicitated or crowd-sourced, and made quantifiable.

	In order to leverage fully the available metadata for a software, and at the same time preserve the reproducibility of the sustainability metric and secure it against manipulation, it is expedient to use more than one metric for the representation of sustainability. E.g., a ``hard'' metric could be computed over metadata that is both quantifiable and objective (such as whether a software is open source or not), a ``semi-hard'' metric over metadata that is objective but hard to quantify (e.g., use of a programming language or build system that in itself may be more or less sustainable), and a ``soft'' metric taking into account qualitative and subjective metadata (e.g., whether a UI is intuitive). Both the ``hard'' and ``semi-hard'' metrics would be reproducible and would likely be hard to tamper with, as they are based on objective metadata. The use of these three metrics would additionally allow for user interaction with the metadata, i.e., users could review and evaluate metadata, and these evaluations could in turn be used to detect bad data and re-compute metrics where necessary. User contributions in the form of, e.g., votes on or grading of specific metadata points would inherently be constituting the ``soft'' metric, while other forms of contributions, such as documenting the use of a software (e.g., in conjunction with a reference to work documenting the use of the software), could contribute to either of the ``harder'' metrics as well. Interactivity can also be used for gamification purposes to attract users, and provide software originators with further incentive to submit their data, e.g., by releasing standings tables and issuing metrics graphics for use on websites, etc.

	The accumulation of the metadata can be performed by different means: direct input by the originator of the software, harvesting data from existing repositories via, e.g., GitHub and Open Hub APIs, dedicated crawling of source code repositories, etc. The latter two methods can also be used for verification and a preliminary quantification of the input. The metadata must be available in the repository as structured text in a well-defined format, e.g., based on XML or JSON.

	In summary, an interactive metadata repository, field-specific or not, that measures and documents the sustainability of software can be a valuable tool not only for the discovery of research software, but also for the identification of sustainable -- and therefore preferable -- software, and as an integral part of the implementation of SMPs.

	In future research, the author intends to further develop the idea of such IMRs within the framework of a PhD thesis.

\bibliographystyle{IEEEtran}
\bibliography{references}

\end{document}